\documentstyle[preprint,aps]{revtex}
\tighten
\begin{document}
\preprint{IFUM 548/FT  March 1997/Revised version}
\draft
\title
 { $N$-vector spin models on the sc and the bcc lattices:\\
a study of the critical behavior of the susceptibility   \\
 and of the correlation length by high temperature series\\
 extended to order $\beta^{21}$\\}
\author{P. Butera\cite{pb} and M. Comi\cite{mc}}
\address
{Istituto Nazionale di Fisica Nucleare\\
Dipartimento di Fisica, Universit\`a di Milano\\
Via Celoria 16, 20133 Milano, Italy}
\maketitle
\begin{abstract}
Abstract: High temperature expansions
for the free energy, the  susceptibility and  
the second correlation moment of
the classical $N$-vector model [also known as the 
$O(N)$ symmetric classical spin Heisenberg  model or as the lattice 
$O(N)$ nonlinear sigma model]
on the simple cubic  and the body centered cubic 
lattices are extended  to   order
$\beta^{21}$ for arbitrary $N$.  
The series for the second field derivative of the susceptibility
 is  extended  to   order $\beta^{17}$. 
We report here on the  analysis of the newly computed series 
for the susceptibility and the (second moment) correlation length
 which yields  updated estimates of the critical parameters
 for various values of the spin dimensionality $N$, including
 $N=0$ [the self-avoiding walk model], $N=1$ [the Ising spin 1/2 model], 
 $N=2$ [the XY model],  $N=3$ [the classical Heisenberg model].
For all values of $N$ we confirm a good agreement with the present
 renormalization group estimates.
A study of the  series for the other observables will appear 
in a forthcoming paper.
\end{abstract}

 \widetext

\section{Introduction}
The continuing interest in the high temperature (HT) expansions 
 for the statistical mechanics of lattice spin 
models  (or equivalently in the 
strong-coupling  expansions for Euclidean lattice 
field theories)
has in the last few years been a strong incentive   
for  a substantial extension 
of series in a variety of models. 
This  valuable computational advance has been 
made possible not only
by  the large improvements of the computers performance and the rapid
growth of their memory capacity in the last decade,
but mainly by a more careful 
reconsideration of well known 
expansion techniques, which have yet to be fully  
exploited, and by a greater effort in 
devising and implementing
 faster algorithms\cite{nota1,bc2d}.  

We have  devoted this paper to  
 the widely studied  $N-$vector model\cite{st68} on 
three-dimensional lattices, 
which is also known as the Heisenberg 
classical $ O(N)$ spin model or, in field theoretic 
language, as the lattice 
$O(N)$ nonlinear sigma model.
 We recall that a     HT expansion of the susceptibility $\chi$
  to order $\beta^{23}$ has  been recently obtained\cite{mac}
  on the simple cubic (sc) lattice 
for $N=0$ [the self-avoiding walk (SAW) model\cite{gen}]
 by a direct walk counting technique which cannot be 
extended to different values of $N$. 
In the $N=1$ case [the Ising spin 1/2  model] 
the series $O(\beta^{21})$  on the body centered cubic (bcc)
 lattice for both $\chi$ and 
the second moment of the correlation function  $\mu_2$, 
 obtained\cite{nickel80,nr90} in the pioneering work by B. G. Nickel, 
remain still unsurpassed, 
but  for $N>1$ the series now available  are significantly shorter. 
 The general situation before our work of which a brief,  
partial and preliminary account  has already 
appeared in Ref.\cite{bc953d}, is summarized in table \ref{tabella1} 
listing the longest   series published until now\cite
{mac,nickel80,nr90,gut89,gaunt,roskies,bcg93,lw88,b90,fere,mckdom,re,r1,sd}   
for specific or generic values of $N$,
on the sc and on the bcc lattices.

This work is part of a sequence devoted to the 
extension and the analysis of HT series 
to order $\beta^{21}$
 for the $N$-vector 
model on $d$-dimensional bipartite  lattices. 
The case of the square lattice has already 
been discussed in Ref.\cite{bc2d}.
 We have chosen to use  the (vertex renormalized) linked cluster
expansion (LCE) technique \cite{lw88,w74mck83,bk,rs} 
because we have developed algorithms which make it  equally efficient
in a wide range of space dimensionalities independently of
  the nature of the site 
 variables, whereas  most other methods give best performances 
only in very specific situations, such as  two-dimensional 
or low coordination number lattices
 and for sufficiently simple interactions. 
 Having thus  avoided  the limitations of previous
 work, we have  been able to produce extensive tables of series  expansion 
coefficients  given as explicit functions of the spin dimensionality $N$, 
  which summarize in a convenient   format 
a large body of information  for an infinite set of universality 
classes. 
 These tables, which we consider as  the main result of our work, are
 reported in the appendices.

We have built upon Ref.\cite{lw88}
where some algorithms  for the automatic LCE calculations
 have been introduced, and HT expansions of 
$\chi$, of  $\mu_2$ and
 of the second field derivative of the susceptibility 
$\chi_4$ for the $N$-vector model on the sc
lattices    have been tabulated\cite{lw88,b90} 
up to $\beta^{14}$ for dimensions $d=2,3,4$. 
   This calculation has been generalized
 to the class of $d$-dimensional bipartite lattices,  
in particular to the (hyper)sc and (hyper)bcc lattices,  
for which  we have  striven to design faster and more 
efficient algorithms and have introduced some  
innovations  dramatically simplifying 
 established  computational schemes.
 In particular,  by   taking full advantage   of the
structural properties of these lattices, 
we have significantly reduced the fast growth
of the combinatorial complexity with the order of expansion which 
 had until now been 
the main obstacle to the extension of the series. 
In fact, it is mainly our effort on the algorithmic side that has made 
 progress  possible, even in comparison with
 more recent work\cite{klom} using 
 the same hardware resources. 
Moreover a considerable 
extension of our  calculations  is still feasible (and it is 
presently ongoing\cite{bcinp}) 
since we are   far from our computational limits.
 Our calculation used an ordinary 
IBM RISC 6000/580H power station 
 with 128 Mb memory capacity and 4 Gb of disk storage. Typical running 
times in the three-dimensional case were a few hours. 
In order to give a rough  idea of the size of the computation 
 it is sufficient to mention that 
over $2\times 10^7$ graphs  have to be generated and evaluated
 to complete the expansion 
of $\chi$ and $\mu_2$ through $\beta^{21}$. 
This should be compared with the  corresponding
figure: $1.1\times10^4$, of  the $O(\beta^{14})$  
computation in Ref.\cite{lw88}
 or with the figure $5\times 10^4$ occurring in the recent analogous 
computation of Ref.\cite{klom} for the face centered cubic 
lattice to order $\beta^{13}$. 
 Approximately $3\times10^6$ graphs
 contribute to the computation of $\chi_4$ at order 17.

We are confident that our results are correct,  
 not only because our codes have  passed numerous 
direct and indirect internal tests, but also because 
$N$ and $d$ enter in the whole computational procedure as parameters, 
so that a good general verification 
is achieved if our expansion coefficients,
 when specialized to $N=$ 0,1,2,3 and $\infty$, 
 agree with the (more or less) long series 
already available in various dimensions.
 Further details on the comparison 
with the available series, in particular in the limits 
 $N \rightarrow 0$ and $N \rightarrow \infty$, can be found in our paper 
 devoted to the two-dimensional $N-$vector model\cite{bc2d}.

Note that, strictly speaking, the $N-$vector model is 
defined only for positive integer spin dimensionality $N$.  
There are, however, infinitely 
many "analytic interpolations" in the variable  
$N$ of the HT coefficients and, 
as a consequence, of all physical quantities. 
 We have performed  the  "natural"
 analytic interpolation of the HT coefficients 
as rational functions of $N$,  
which coincides with that used in the $1/N$ expansion as well as in the
usual Renormalization Group (RG) treatments and 
is  unique 
in the sense of the Carlson theorem\cite{boas}.

It is also worth emphasizing that  the LCE 
technique can be readily adapted
to produce HT expansions for the very general class of models 
(which include the O($N$) symmetric 
$P(\vec \varphi^{ 2})$ lattice boson field theories),
described by the partition function
 
\begin{equation}
Z= \int \Pi d\mu(\vec \varphi_i^{ 2})
exp[ \beta \sum_{\langle r,s \rangle}\vec \varphi_r \cdot \vec \varphi_s],
\label{eq:uno}
\end{equation}
 
 where $ \vec \varphi_i$ is a $N$-component vector 
and $d\mu(\vec \varphi_i^{ 2})$ is the appropriate single spin measure.
If we choose the form 
$d\mu(\vec \varphi_i^{ 2}) = \delta(\vec \varphi_i^{ 2}-1)d\vec \varphi_i$
 for  the single spin measure, 
(\ref{eq:uno}) reduces to the partition function of
  the $N$-vector model. Also 
 a broad class of other
models of interest in Statistical Mechanics 
including the general spin $S$ Ising model, the Blume-Capel model,  
the double Gaussian model,  etc.
can  be represented in this form.
The HT series for some of these models have  been extended and
 will be  discussed elsewhere\cite{bcinp}.
A wider discussion 
of (\ref{eq:uno}) as a full theoretical laboratory 
for the study of scalar isovector lattice 
field theories as well as a detailed account of the 
 graph theoretical and the algorithmic  part of 
our work will  also  be presented elsewhere\cite{bcinp}.

 The  general interest in a  {\it direct}
determination of the critical properties of the classical 
lattice spin models with 
increasing reliability
is clear. Other good general motivations 
 for such a laborious
  calculation as a long HT series expansion include
 more accurate tests  of the validity both  of the assumption of
universality,
 on which the Renormalization Group (RG)
approach to critical phenomena is based, and of
the various approximation procedures
required to   estimate  universal critical parameters
by field theoretic methods.
In fact, waiting for  rigorous arguments to come,
 the only  crucial tests\cite{zinn}  of the validity
of the Borel resummed $\epsilon= 4-d $ expansions\cite{zinn,epsi,epv}  
or  of the perturbative expansions at fixed dimension (FD)
\cite{zinn,epv,zib,mur,ant},
 for which the $O(N)$ model  served as a paradigm, 
are presently limited   to a careful comparison
with experimental data or with other numerical data.
Actually, this comparison  mainly concerns  numerical 
data of different origins, both because
experiments are difficult and the experimental 
results, when available, 
are  much less precise [with the very remarkable  exception of 
the exponent $\nu$ in the $N=2$ case\cite{ahl}] and 
because   experimental 
representatives are known only for a few 
universality classes\cite{mukamel}. 
 Actually  for $N \gtrsim 3$ the physical, but not the numerical, 
interest of the model is somewhat lowered by the observation that the  
$O(N)$ symmetric fixed point appears to be unstable within the 
$\epsilon$-expansion\cite{kl}.

 It should also  be observed   that, 
in spite of  steady progress \cite{wolff}, the stochastic algorithms
 do not yet  seem  ready to completely supersede  HT series
 in the study 
of models where the 
site variables have many components and/or the space dimensionality
is large, or in the computation of	
multispin correlation functions. More  generally HT series 
 remain  valuable  subjects  of independent
study and sources   of auxiliary information 
 for other kinds of 
numerical calculations.   
Therefore we  have also reported in the tables 
our estimates of  nonuniversal critical parameters like 
the inverse critical temperatures.
 The computation of the  critical amplitudes and of 
their universal combinations\cite{aha} 
 will be discussed elsewhere\cite{bcinp}.

The paper is organized as follows: In section 
2 we present our notation and 
 define the quantities that we shall study.
 The analysis of the series is presented in section 3
along with a comparison to some previous analyses, to some 
results obtained by  stochastic methods and 
to the RG results, both by 
the $\epsilon$-expansion and the FD perturbative  techniques.
In the appendices we have reported  the closed form expressions for the
 HT  series coefficients of $\chi$ and $\mu_2$ 
as functions of the spin dimensionality $N$
and their evaluation for  $N=0$ [the  SAW model], 
$N=1$  [the Ising spin 1/2 model], 
 $N=2$ [the XY model] and $N=3$ [the classical Heisenberg model].
The present tabulation extends significantly and supersedes 
the one to order $\beta^{14}$ in Ref.\cite{b90} 
which, unfortunately, contains a few misprints.

In a forthcoming  paper\cite{bcinp} we will present an analysis of
the series for the free energy and for its fourth 
field derivative $\chi_{4}$.

\section{ Definitions and notations}

For convenience of the reader we list here 
our definitions and notations.
As the Hamiltonian $H$ of the $N$-vector model we shall take:

\begin{equation}
H \{ v \} = -{\frac {1} {2}} \sum_{\langle \vec x,{\vec x}' \rangle } 
v(\vec x) \cdot v({\vec x}').
\label{hamilt} \end{equation}

where $v(\vec x)$ is a  $N$-component classical spin 
of unit length at the lattice site $\vec x$,   
and the sum extends to all nearest neighbor pairs of  sites.

The susceptibility is defined as

\begin{equation}
\chi(N,\beta) = \sum_{\vec x} \langle v(0) \cdot v(\vec x) \rangle_c = 
 1+ \sum_{r=1}^\infty a_r(N) \beta^r  \label{chi} \end{equation}
 
where $\langle v(0) \cdot v(\vec x) \rangle_c$ is 
the connected correlation function between the spin at 
the origin and the spin at the site $\vec x$.

The second  moment of the correlation function is defined as

\begin{equation}
 \mu_{2}(N,\beta)=\sum_{\vec x} \vec x^2 \langle v(0) \cdot v(\vec x) 
\rangle_c = 
 \sum_{r=1}^\infty s_r(N) \beta^r. 
\end{equation}

In terms of $\chi$ and $\mu_{2}$ we define the 
 second moment correlation length $\xi$ by 

\begin{equation}
 \xi^{2}(N,\beta)= \frac  {\mu_{2}(N,\beta)} {6\chi(N,\beta) }. 
\end{equation}

\section{ Analysis of the series} 

Let us now turn to a discussion of
 our updated estimates  for the
critical temperatures and the critical exponents
$\gamma$ and $\nu$ in the $N=2,3,4$ cases   
where our new series
are significantly longer  (up to 10 more terms) than those
previously available. We shall also give some comments on the 
 cases $N=0$ and $1$ in which our extension is more modest and 
for $N > 4$ where only a few numerical results are available.

 It  has become clear from a long experience,  
 mainly gained from the analysis of the Ising 
model HT expansions\cite{nickel80,gaunt80,zinn79,fisher,adl,george}, that  
in order to achieve 
a substantial improvement in the precision of the estimates 
of the critical parameters  
from the analysis  of extended series 
 one should properly allow for the  expected
nonanalytic  corrections\cite{wegner} [usually also called 
 confluent corrections] to
the leading power
law behavior of thermodynamic quantities near a critical point. 
For instance, we recall that, if we set $\tau= 1- \beta/\beta_c$,
   the susceptibility is expected to behave, in the 
vicinity of the critical point $\beta_c$, as
 
\FL
\begin{equation}
 \chi(N,\beta)
\simeq C_{\chi}(N)\tau^{-\gamma(N)}\Big(1+ a_{\chi}(N)
\tau^{\theta(N)} + a'_{\chi}(N)
\tau^{2\theta(N)}+...
+ e_{\chi}(N)\tau + e'_{\chi}(N)\tau^2+...\Big)
\label{conf}\end{equation} 

when $ \tau \downarrow 0$.
 Not only the critical exponent $\gamma(N)$, but also
 the leading confluent  correction 
exponent $\theta(N)$ is universal (for each $N$).
On the other hand, the critical amplitudes $C_{\chi}(N), a_{\chi}(N), 
a'_{\chi}(N), e_{\chi}(N),$ etc. are expected to
 depend smoothly on the parameters of the Hamiltonian, i.e. 
they are nonuniversal.
 Similar considerations apply to $\xi$ 
(and to the other singular quantities)
 which, however, contains a different critical
 exponent and different critical amplitudes 
$C_{\xi}(N), a_{\xi}(N),$ etc., but
the same leading confluent exponent $\theta(N)$.
 It is also known that  $\theta(N) \simeq 0.5$
 for small values of  $N$\cite{zinn}
and $\theta(N)=  1 +O(1/N)$ for large $N$\cite{ma}.

As   experience has indicated, the established ratio  extrapolation
 and Pad\'e approximant (PA)  methods are generally inadequate to 
the difficult numerical problem of determining 
 simultaneously $\beta_c$, the critical exponent 
 and the leading confluent exponent in Eq. (\ref{conf}), 
 a task which essentially amounts to an  intrinsically  unstable  
double exponential fit.
It is considered appropriate, then, to resort
 to the  inhomogeneous  differential
 approximants (DA) method \cite{gutasy}, a
generalization of the PA method, which,  in principle,  can be better
suited to represent functions behaving 
like $\phi_1(x) (x-x_0)^{-\gamma} +
\phi_2(x)$ near a singular point $x_0$, where $\phi_1(x)$
is a   regular function of $x$  and $\phi_2(x)$ may 
contain a (confluent)
singularity of
strength smaller than $\gamma$.

\subsection{Unbiased analysis}

 To begin 
with, we have performed 
 a  series analysis by DA's, 
 essentially following the protocol 
 suggested in Ref.\cite{gut87} which is {\it unbiased} for
confluent singularities. For each $N$,  
we have computed $\beta_c(N)$ and $\gamma(N)$  
by first and second order DA's 
built in terms of the susceptibility series and have 
then used this estimate of $\beta_c(N)$
 to bias the determination of $\nu(N)$ from the series for
 the square of the (second moment) correlation length $\xi^2$.
Completely consistent results are also obtained, in general, 
by the method of critical point 
renormalization\cite{gutasy}.
Also the specific heat exponent $\alpha(N)$ can be 
estimated by examining
the behavior of $\chi$ at $\beta=-\beta_c(N)$,  where a 
 weak antiferromagnetic singularity is expected for bipartite lattices.
As shown  in Ref.\cite{fiaf}, having set $\tilde\tau = 1 +\beta/\beta_c$, 
 one has  

\begin{equation}
 \chi(N,\beta)
\simeq \tilde c(N) + \tilde a(N) \tilde \tau^{1-\alpha(N)} 
+ \tilde b(N)\tilde\tau +...
\label{saf}\end{equation} 

 as $\tilde \tau \downarrow 0$. We shall however present this study 
in  a forthcoming paper\cite{bcinp}
  in order to jointly discuss  also the results of the 
analysis of the free energy.

The  results  of the present unbiased analysis  
do not significantly modify 
those obtained in the similar preliminary 
study\cite{bc953d} with series $O(\beta^{19})$.
They are reported in 
table \ref{tabella2} for $N\leq 3$ and in table \ref{tabella3}
   for $N \geq 4$ and compared with some of the most 
accurate recently published  estimates 
\cite{mac,epv,zib,mur,ant,ahl,lms,ca,blh,adh,has,jan,che,bal,fer,kan,mal,oka}
by various other methods, in particular by RG perturbative methods.
 Only for the physically most interesting cases, namely for
 $N=0,1,2,3$, elaborate Borel resummed 
estimates are available for both the fifth order 
$\epsilon$-expansion\cite{zinn,epsi,epv} 
 and either the six loop\cite{zib} or the seven loop FD expansion\cite{mur}.

 The scope of the seven loop 
 FD computation of  
Ref.\cite{mur}  is however slightly limited 
by the present uncertainty in the value of the renormalized coupling
constant $\bar g(N)$ used  in the calculation. 
Therefore, in Ref.\cite{mur}, the exponent values  have 
been conveniently  expressed as the  sum 
of the  central estimate corresponding to this approximate 
value  and of  a small deviation 
 proportional to the difference between $\bar g(N)$ and the true 
 renormalized coupling $ g^*(N)$. 
For simplicity, we have reported in table \ref{tabella2} 
the central estimate and have allowed for the  contribution 
of the possible deviation  only by doubling the 
expected error of the summation 
procedure: this roughly amounts to assume optimistically an uncertainty in
 the value of the renormalized coupling of a few parts per thousand. 
 For $N > 3$, no estimates  of the 
exponents by the $\epsilon-$expansion 
method have been published, while only 
very recently an extensive computation by 
the six loop FD expansion method has 
appeared \cite{ant}. 
Unfortunately, no estimates of error for 
the exponents are given 
in Ref.\cite{ant}, but, in analogy with the small $N$ case,  we 
would reasonably expect relative errors of the order of $1 \%$.

From  our unbiased analysis of the sc lattice series we obtain  exponent 
estimates essentially consistent, within their errors, with the available 
$\epsilon-$expansion results, 
but, in general, slightly larger (up to  $\simeq 1\%$, or even more 
for intermediate values of  $ N $) than the 
 FD expansion results.  

In the case of the bcc lattice, 
however,  our unbiased estimates are  also  completely compatible  
with the sixth order \cite{zinn,zib} FD perturbative results 
or with the most recent seventh order
\cite{mur} results.
A similar situation has already been  encountered  in a previous
very accurate unbiased analysis of the $N=1$ case\cite{gut87}.
 We see this simply as a confirmation  that the series
for lattices with larger coordination number
have a  faster convergence rate 
[or, in other words, a greater  "effective length"\cite{gut87}] and 
 also as an indication
 that   the simplest unbiased DA's might be only 
partially able to describe the  confluent singularities.
 Therefore the larger discrepancies of the sc lattice results 
should not be interpreted as 
indicative of universality violations, 
 but rather as a warning that  
the systematic errors of our analysis 
due to  the finite length of our series and to the 
confluent singularities are quite likely to be 
underestimated if we evaluate the uncertainties solely
 from the scatter of the approximant values obtained using a 
sufficiently large number of series coefficients.  We should also 
 stress that,  for $N \geq 4$, the sequence of DA estimates 
for the critical temperature or the exponents which use an 
increasing number of coefficients, show  evident
residual  trends  which indicate the presence of important 
 confluent corrections to scaling, so that some
 "reasonable"  extrapolation of the results becomes necessary. 
Since this  inevitably 
involves some assumption on the confluent 
exponent and therefore introduces  
 some biasing, it will be more appropriately 
dealt with in the next paragraph. In order to distinguish clearly 
the effects of the various assumptions, we have chosen not to  perform
 any further extrapolation in our
 "unbiased estimates" reported  
in tables \ref{tabella2} and \ref{tabella3},
 although  it is  clear that  neglecting residual trends 
is a  source of sizable systematic error. 

Even within these limitations, we have 
improved the precision of the values of the critical parameters
as obtained from HT series by unbiased methods, and, so far,   
have not  inferred from this analysis any  indication of a 
serious inconsistency with the estimates  from RG.

In conclusion, within the {\it unbiased} approach to series analysis, 
the influence of the confluent singularities 
 can be assessed more accurately  and the results of 
the   analysis can be better reconciled with the estimates from
the RG methods probably only by  computing 
 still longer series, as it has been 
already recognized for the Ising model\cite{nickel80,zinn79,fisher}.
On the other hand, if  we are ready to assume universality
 from the beginning, then, as 
suggested  by work on the bcc lattice in the $N=1$ case\cite{nr90,fisher} 
whose results are reported in table \ref{tabella2},  
a more accurate determination 
of the critical exponents could be obtained,  even without 
further extending the series, 
 by a study of appropriately built    families of
models  depending on some continuous parameter,
    for each universality class.
 The idea is, essentially, to minimize or suppress the 
 amplitude of the dominant confluent correction 
to scaling in Eq. (\ref{conf}) by taking advantage 
 of its nonuniversality, namely of its continuous dependence 
on the parameter entering in the Hamiltonian.
 Using our LCE computation, it is now possible to implement  easily
this procedure for any $N$ and on two different lattices, 
 thereby  corroborating its reliability. These developments of our study
 and a  detailed analysis of other features of the series 
including  estimates of the universal ratios 
of the leading confluent amplitudes for
 $\chi$ and $\xi^2$ will be presented elsewhere\cite{bcinp}.

\subsection{Biased analysis}

We have also analyzed our series by various  {\it biased} methods, 
 in particular  by 
 using  properly designed
 first order inhomogeneous DA's  in which both $\beta_c(N)$ and
  the correction to scaling exponent $\theta(N)$  have been
fixed , or
   by second order inhomogeneous  DA's 
 in which   only $\theta(N)$\cite{nr90,nick81}   has been fixed.

In order to provide the additional information 
needed in these approaches, 
for $N\leq3$, we have assumed  that the exponents  $\theta(N)$ take the 
values predicted by the FD perturbative RG \cite{zinn}  

\begin{equation}
\theta(0)=0.470(25),   \theta(1)=0.498(20), 
 \theta(2)=0.522(18),  \theta(3)=0.550(16).
\label{expth1}\end{equation}

For $N>3$,  the first FD perturbative 
estimates at six loop order  have only been obtained  very 
recently by A. I. Sokolov \cite{sokoth} and 
 have been kindly communicated to us before publication
\FL
\begin{eqnarray}
\theta(4)=0.578(10),  \theta(6)=0.626(10), \theta(8)=0.670(10), 
\theta(10)=0.707(10), \theta(12)=0.737(10).
\label{expth2}\end{eqnarray}

We have therefore been able to revise the biased analysis 
presented in the first  preprint version of this paper, where, 
for lack of a better choice, we had used  values for $\theta(N)$ ($N>3$) 
(larger than those in Eq. (\ref{expth2})), 
obtained by a reasonable, but unwarranted interpolation method.
   In some cases, in particular for $N=0$ in Refs.\cite{mac,lms,ishi} 
and for $N=1$ in Refs.\cite{nr90,zinnfish},  
 numerical work has  suggested
 slightly different  confluent exponents, 
 which might be more accurate or, to some extent, 
 also provide an "effective" description  of higher confluent
corrections. 
We shall return later to this point, but we should mention that, 
at the level of precision of the following calculations,  
 the precise values 
of the  exponents $\theta(N)$ do not matter too much, since 
we have observed 
 that our biased estimates of the leading critical exponents 
  remain practically stable (within their errors) 
 under   variations  of the confluent exponents up to 
$\simeq 5-10\%$. Of course other quantities, such as 
the confluent amplitudes,  are very sensitive to the 
values of $\theta(N)$.

Let us now sketch the first biased method.
If an  accurate estimate for $\beta_c$ is available, we can formulate a
quite simple procedure, biased with  both 
$\beta_c$ and $\theta$.
From the asymptotic formula (\ref{conf}) we get
\begin{equation} 
 \gamma(N,\beta) \equiv (\beta_c(N)-\beta) 
\frac {d{\rm log}\chi(N,\beta)} {d\beta} =
 \gamma(N)+ \rho(N)( \beta_c(N)-\beta)^{\theta(N)} 
+ o\big( ( \beta_c(N)-\beta)^{\theta(N)}\big) 
\label{gamma}\end{equation} 
where $\rho(N)=-\theta(N) a_{\chi}(N)$.
 
We can approximate the quantity $\gamma(N,\beta) $
 by the particular class of first order inhomogeneous  DA's 
defined as the solutions of the equation
\begin{equation} 
Q_m(\beta)\Big[(\beta_c(N)-\beta)\frac {dy}{d\beta}+
 \theta(N) y(\beta)\Big]
+R_n(\beta)=0
\label{damod}\end{equation} 

Here $Q_m(\beta)$ and $R_n(\beta)$ are  polynomials of degrees $m$ and $n$
 respectively, calculated, as usual,
 from the known series expansion of $\gamma(N,\beta)$.
 As a result the  exponent $\gamma$ is simply estimated as 
\[\gamma(N)_{m,n} = \frac {-R_n(\beta_c(N))} {\theta(N) Q_m(\beta_c(N))}\]
and the amplitude $\rho(N)$ of the subleading term in Eq. (\ref{gamma}) is 
given  by the formula
\begin{equation} 
 \rho(N)_{m,n}=\frac {y(0)-\gamma(N)_{m,n}}{{\beta_c(N)}^{\theta(N)}}
   - \int_0^{\beta_c(N)}\frac { D(t)_{m,n}dt}
{ {(\beta_c(N)-t)}^{1+\theta(N)} } 
 \label{amps} \end{equation} 

where

\[D(t)_{m,n}= \frac{R_n(t)} {Q_m(t)} - 
\frac {R_n(\beta_c(N))} { Q_m(\beta_c(N))}\]

We  consider only almost diagonal approximants, namely those with
 $\vert m-n \vert \le 4$.
This procedure,  which might be seen as the simplest, although
 certainly not the most general, DA 
extension of the biased PA 
method introduced in Refs.\cite{rosk,adp},
works accurately on model series having the analytic structure 
expected for $\gamma(N,\beta) $ in the vicinity of $\beta_c(N)$.
A similar procedure can be applied to $\xi^2(N,\beta)/\beta$ 
in order to compute the exponent $\nu(N)$.
In order to give an idea of the results, let us for example, 
 consider  the $N=1$ sc lattice series. 
Assuming $\beta_c =0.2216544(3)$ as suggested in Ref.\cite{blh}
 and $\theta=0.498(20)$,  
we estimate $\gamma=1.2388(10)$ and $ \nu=0.6315(8)$.
 With the same value of $\theta$, in the  bcc lattice case, assuming 
$\beta_c =0.157373(2)$ as suggested in Ref.\cite{nr90},
we obtain $\gamma=1.2384(6)$ and $ \nu=0.6308(5)$.
 In general, when the value of $\beta_c$ is not known accurately, 
  it would be more appropriate 
to present the results  in the form
 of a linear relationship between  the critical exponent and $\beta_c$, 
at  a given fixed value of $\theta$.
 We can, however,  use this method also to determine the 
value of $\beta_c$ by fixing 
only $\theta$ and looking  for the (generally small) range of values of 
$\beta_c$ for which the uncertainty of the exponent $\gamma$  is minimal.
The estimates obtained in this way are in general completely consistent
with those from the analogous biased PA procedure of Refs.\cite{rosk,adp}.
The values of $\beta_c$ so obtained  are also, in most cases, consistent
with those obtained  
from the unbiased improved ratio methods  of Ref.\cite{zinn79}
after extrapolating the sequences of results linearly
in $1/r^{1+\theta(N)}$ (where $r$ is the number of series 
coefficients used), and with the estimates obtained 
 by similarly extrapolating  the results from unbiased DA's.
Analogous considerations apply to our exponent estimates, which, however, 
 have to  be compared to the results obtained from improved ratio methods 
by extrapolating linearly in $1/r^{\theta(N)}$.

 A  second method,  modelled after Refs.\cite{nr90,nick81},
 which is biased only with the value of the 
confluent exponent, may be described as follows:
 for each value of $N$, we   have considered 
the approximants  derived from inhomogeneous second 
order  differential equations  with the structure
\[ (\beta_c(N)-\beta)^2(\beta_c(N)+\beta)Q_l(\beta) 
\frac {d^2y} {d\beta^2} 
 +(\beta_c(N)-\beta) P_m(\beta) \frac {dy} {d\beta} 
+ R_n(\beta) y(\beta) + T_s(\beta)=0\]
 where $Q_l(\beta), P_m(\beta), R_n(\beta)$
 and  $T_s(\beta)$ are polynomials of degrees $l,m,n$ and $s$ in 
 the variable $\beta$.
 By this choice the DA's  are biased
to be singular at $\beta=\beta_c(N)$ and 
at $\beta=-\beta_c(N)$.
 We restrict ourselves to almost diagonal 
DA's (namely those with $l+3 \simeq m+1 \simeq n$ and $s\leq 4$), 
  which use at least 19 series coefficients. 
For each DA, we  adjust 
$\beta_c(N)$ in a small range around the values indicated 
by the  unbiased analysis of the previous section until the correction 
to scaling exponent $\theta(N)$ reaches precisely the central value 
indicated in Eqs. (\ref{expth1}),(\ref{expth2}). 
 The corresponding inverse critical 
temperature and exponents 
are then taken as the best estimates of these quantities. 
 It should  be noticed that,  within this approach, 
the values of $\beta_c$ for $\chi$ and
 $\xi^2$ cannot be forced to be equal,
 but the differences are generally 
not much larger than the errors.
In this approach (like in the previous one)
 the evaluation of the errors  is based,  as  usual,
   on the spread among the approximant values and includes  a  
 generous allowance for 
the uncertainty in  the biased value of $\theta(N)$.
 Let us finally mention that
we also have  preliminarily tested 
the reliability  of this procedure on various
  model series built in such a way to reproduce the main 
expected features of $\chi$  and $\xi^2$.

An important remark concerns the values of the amplitudes of the 
confluent corrections in $\chi$ and $\xi^2$, which can be obtained 
from Eq. (\ref{amps}).
We have observed that, in the bcc
lattice case,
 they are 
negative for $N \lesssim 2$,  positive  and  increasing 
for $N >2$. A quite similar behavior of these  amplitudes
 is observed in the sc lattice case,
 where the change in sign occurs for $N \simeq 3$.
  For $2<N<4$,  the amplitudes are relatively small for both lattices, 
 but they become important for $N>4$. 
This remark, which will be illustrated in full 
 detail elsewhere\cite{bcinp}, is  also consistent with the fact that 
our  biased estimates of the critical exponents 
 are increasing functions of the confluent
 exponents for $N\lesssim 2$, while  
 they are decreasing functions for $N\gtrsim 4$.
 For $ 2 \lesssim N \lesssim 4$ some approximants give 
increasing functions and others give decreasing functions.
 Completely consistent  indications on the behavior of the confluent 
amplitudes can also be inferred 
 from the features of the improved ratio plots\cite{zinn79}, or 
by studying, as function of $N$, the difference among the unbiased 
DA  estimates of $\beta_c$ 
 obtained from $\chi$, $\mu_2$ or $\xi^2$, 
as suggested in Ref.\cite{fisher}.
In  the $N=1$ case, our remark agrees with the results of 
earlier studies\cite{nr90,fisher,liufi,george} 
 on the sc, bcc and fcc lattices, which established that the sign of the
 leading confluent amplitudes in $\chi$ and $\xi^2$ is negative.

Some features of the biased procedures we have adopted,  may appear
questionable or  may call for further improvement. We have always tried 
to minimize the possible defects by 
forming our final estimates with the, generally quite compatible,
indications coming from all the various available biased methods 
and not only from the two biased DA procedures we have described above.
 Moreover we have  indicated very conservative error bars. 
It is  however  reasonable to expect that  
the accuracy of the biased estimates
  depends not only on whether the present series are long enough 
to provide  the essential information on the subdominant singularities, 
but also on whether most of the corrections to scaling 
can actually be well described  by the first nonanalytic 
term in the asymptotic  formula (\ref{conf}). 
  We believe, however, that the final results are at least 
consistent and  suggestive.
First of all, our  analysis confirms that the total  
(statistical + systematical) errors of 
the previous unbiased approach are   larger than we have indicated.
 For $N\leq 1$ the biased exponent estimates differ only slightly
from the unbiased ones and in such a way to improve the
 agreement with the most accurate  FD perturbative  
estimates, with stochastic simulations and with 
experimental results. For $N=2,3$ the agreement with the RG, in the 
 FD perturbative approach, is, perhaps, less convincing.
For  $N \geq 4$ our biased exponent estimates are
systematically larger (up to $\simeq 4\%$) than the FD six loop 
perturbative values\cite{ant}.
 This discrepancy 
is parallel to and, in our opinion, is  
strictly related to the fast increase of the confluent 
amplitudes of $\chi$ and $\xi^2$
in the same range of $N$.  Therefore the problem 
seems to be of a purely numerical nature. 
Of course, we cannot prove that, in these circumstances,  
our biased analyses can allow for 
 the confluent singularities better than 
the present Pad\'e-Borel resummation
 techniques do within the FD perturbative approach.
We can perhaps only argue that, at least for large $N$, our results might
 be more accurate by comparing them with 
the results of  the $1/N$ expansion of the exponents.
 Indeed for $N \gtrsim 14 $, the $1/N$ estimates,  
 which seem  to be rather  
well converged  and therefore reasonably accurate, 
  seem to approach  our biased bcc estimates
 faster than the FD values. 
Let us finally add that these numerical problems with the perturbative 
estimates of the exponents 
might also be partly due to possible  residual imprecisions 
in the values of the renormalized coupling constants entering 
in the FD perturbative calculation\cite{mur}. We shall return to 
this point when new 
estimates of the coupling constant will be available 
from HT series\cite{bcinp}.

\subsection{Final comments}

Let us now finally add some comments on the  results 
of the series analysis which are presented  in the  tables.

In the $N=0$ case, the SAW model,  we have not attempted either to 
report in  table \ref{tabella2} or to cite in the references 
 even only a representative sample of 
 the large amount of numerical work accumulated over the years, 
which fortunately  has  been reviewed recently in the very extensive 
 and valuable new
treatise\cite{hughes} and in Ref.\cite{lms}
devoted to   a high  precision stochastic  study.

On the sc lattice, our series for $\chi$ and $\mu_2$ 
are not yet longer than those of Ref.\cite{mac} and of Ref.\cite{gut89} 
respectively, but we have  taken
 advantage of the two additional published expansion coefficients 
of $\chi$  in order to test the stability of our biased estimates.
 We recall that the previous HT analysis by 
unbiased DA's of the  sc lattice 
series to order $\beta^{21}$ performed in Ref.\cite{gut89}  produced 
the estimates $\beta^{sc}_c=0.213496(4)$, $\gamma= 1.161(1)$ and 
$\nu=0.592(4)$, which are all completely consistent 
with the results of the later analysis of Ref.\cite{mac} and with 
our own unbiased analysis, but slightly larger than our biased 
estimates. Very recently the significantly lower estimates 
$ \beta^{sc}_c=0.213492(1)$ and $\gamma= 1.1575(6)$ have been 
obtained by a stochastic method sampling 
SAW's which extend up to $ 4\times 10^4$ steps\cite{ca}.
 Using this estimate of $\beta^{sc}_c$ in  the first  DA method 
introduced in the previous section, we get $\gamma=1.1589(8)$; 
 in order to obtain the lower value $\gamma \simeq 1.1575$ 
 a value $\beta^{sc}_c \simeq 0.213488$ would be required, 
 which  is  somewhat below the range suggested by the 
presently available HT series on the sc 
lattice, even allowing for the confluent 
corrections. 

In the bcc lattice case we have computed five new
coefficients for $\chi$ and $\mu_2$ beyond 
those reported in Ref.\cite{gut89}. 
 This makes it worth computing new  estimates for  the exponents.
 We  recall that the  analysis in Ref.\cite{gut89} of the
$O(\beta^{16})$ bcc lattice series  available  until now yielded the 
values $\beta^{bcc}_c=0.153137(10)$, $\gamma= 1.162(2)$ and 
$\nu=0.592(2)$, which  are 
 less precise, but compatible with our new unbiased estimates and somewhat
larger than our corresponding biased estimates.
On the other hand, our biased estimate of $\gamma$ for the bcc lattice 
 agrees more closely with the new stochastic estimate\cite{ca}. 
We should also stress that, for both the sc and the bcc lattices,  
 our biased estimates of $\nu$    have now come very close to 
the RG estimates and to the 
experimental value $\nu=0.586(4)$ reported in Ref.\cite{cot}.

  The value of $\theta(0)$
 is still controversial\cite{ishi}. For example,  the study\cite{mac}  
of the long HT series
available on the sc lattices suggests $\theta(0)\simeq 1.$, 
 while an extensive MC study\cite{lms} on the same lattice 
 rather indicates  an effective exponent  
$\theta(0)\simeq 0.56(3)$.
  Assuming this last value in our computation instead of the 
one in Eq. (\ref{expth1}), would only 
shift the  biased  estimates of $\gamma$ and $\nu$, 
in the bcc lattice case,
 from 1.1582(8) to 1.1587(8)  and from 0.5879(6) to 0.5883(6)
respectively. Similarly in the sc
 lattice case  the estimate of $\gamma$ (using all 23 available 
coefficients) would change from
 1.1594(8)  to 1.1597(8) and that of $\nu$   from 0.5878(6) to 0.5894(8).

A final remark is  that 
from our extended series for $\chi$ on the bcc 
lattice we can derive\cite{ham}  the new rigorous and
 stricter    inequality
$\beta^{bcc}_c(0) \geq 
exp(-\frac{1} {21} {\it ln} a_{21}(0)) \simeq 0.148582..$, 
 which slightly improves the previous bound obtained from  $a_{16}(0)$ 
and quoted in Ref.\cite{hughes}.

In the $N=1$ case, the Ising spin 1/2 model, the  relevant 
 numerical studies  are even more numerous than for the SAW model,  
so that we can only 
address the reader to the recent extensive 
 review which in  Ref.\cite{blh} 
complements  stochastic computations  of unprecedented 
accuracy on the sc lattice.
 On this lattice, we have extended by two terms 
the $\chi$ series and
by six terms the  $\mu_2$ series\cite{roskies}   
 so that it is worthwhile at least to update the  estimate of $\nu$.
In order to compute the exponents reported in table \ref{tabella2}    
from the sc series, we have  
assumed  $\theta(1)=0.498(20)$ and simply taken 
the extremely accurate value 
$\beta^{sc}_c(1)= 0.2216544(3)$ 
obtained in Ref.\cite{blh} and therefore indicated within parentheses. 
Our bcc lattice series for $\chi$ and $\mu_2$  
 are  not yet longer than 
the  series of Refs.\cite{nickel80,nr90}. 
 In this case, we have assumed the same 
value of $\theta(1)$ and taken 
$\beta^{bcc}_c(1)= 0.157373(2)$ from Ref.\cite{nr90} (similarly 
 indicating it within parentheses).
Also for the Ising  model, values of the confluent exponent 
 slightly larger than  the one we have assumed,
 such as  $\theta(1)=0.52(3)$ in Ref.\cite{nr90} 
and $\theta(1)=0.54(3)$ in 
Refs.\cite{fisher,zinnfish} have been reported.
 Taking the largest of these  values in our 
biased computation would 
 only change,  in the bcc lattice case, 
the central estimate of $\gamma$  from $1.2384$ to $1.2387$ 
 and would shift that of  $\nu$ from $0.6308$ to $0.6311$.  In the sc 
lattice case the central value of $\gamma$
 would be shifted from $1.2388$ to $1.2392$ and that of $\nu$ from $0.6315$  
 to $0.6321$.

In the $N=2$ case, the XY model,  we have computed four more terms of the 
$\chi$ and $\mu_2$ series 
in the sc lattice case. 
Notice that the last two coefficients of the 
 previously published   $O(\beta^{17})$ 
series contained tiny 
numerical errors (inconsequential for the analysis in Ref.\cite{bcg93}) 
 which are corrected  by our new computation.
 In the case of the bcc lattice, our extension of the series 
 for $\chi$ and $\mu_2$ amounts to nine 
terms and gives a greater 
significance to the new exponent estimates. 

In the $N=3$ case, the classical Heisenberg model,  we have 
extended by seven terms the series for $\chi$ and $\mu_2$ 
on the sc lattice.   
 In the case of the bcc lattice we have extended the series 
by ten terms.

For all $N>3$, only series  up to $O(\beta^9)$
 were available until now on the bcc lattice and therefore our extension 
amounts to twelve terms.
 On  the sc lattice 
 we have computed seven additional series 
coefficients for $\chi$ and $\mu_2$.
Let us finally note that, on the sc lattice for $N=2,3$ and 
$4$, the estimates of 
$\beta_c$ indicated by the simulation of Ref.\cite{bal} are only slightly
smaller than ours. Using these values in our biased DA method, we would
 obtain  $\gamma \simeq 1.322$, $\gamma \simeq 1.402$ and 
$\gamma \simeq 1.476$, for $N= 2,3$ and $4$, respectively. For the exponent
 $\nu$ the agreement would be somewhat closer.      

\section{ Conclusions} 

In conclusion, we have produced new HT expansions through order 
$\beta^{21}$ 
of the susceptibility and of the second correlation 
moment for the classical
$ N$-vector model with general $N$, on the sc and the bcc lattices. 
This rich material has been conveniently tabulated in the 
appendices in order to offer an easy opportunity for further study.

As a first application of 
our results, we have updated the direct estimates of the critical 
parameters of the $N$-vector model with
 a considerable improvement 
in  accuracy over previous analyses  and, for all values of   $N$,
we have confirmed a generally good agreement
 with the most precise calculations by current approximate RG methods.

\acknowledgments 
This work has been partially supported by MURST. We thank 
Professor A. J. Guttmann
for critically reading the first draft of this paper and
 Professor A. I. Sokolov for making available to us before 
publication his very useful results on the perturbative computation 
of the confluent exponents for $N>3$.

\begin{table}
\caption{ Longest published  HT expansions 
 for the $N$-vector model
 on the simple cubic  and the bcc lattice before this work.}
\label{tabella1}
\begin{tabular}{rrrrr}
 &  Quantities expanded &  Parameters &  Maximal order & Ref.\\
\tableline
sc lattice&  &   & &    \\
*&   $\chi$&           $N=0$&         23&\cite{mac}\\
*&   $\chi,\mu_2$&     $N=0$&         21& \cite{gut89} 	\\
*&   $\chi$&           $N=1$&         19& \cite{gaunt} 	\\
*&   $\mu_2$&          $N=1$&         15&  \cite{roskies}	\\
*&   $\chi,\mu_2$&     $N=2$&         17& \cite{bcg93}	 \\
*&   $\chi,\mu_2$&   any $N$&         14&	\cite{lw88,b90} \\
\tableline
bcc lattice&  &   & &    \\
*&   $\chi,\mu_2$&     $N=0$&           16&	\cite{gut89}\\
*&   $\chi,\mu_2$&     $N=1$&           21&	\cite{nickel80,nr90}\\
*&   $\chi,\mu_2$&     $N=2$&           12&	\cite{gut89}\\
*&   $\chi$&           $N=3$&           11&	\cite{mckdom}\\
*&   $\chi,\mu_2$&   any $N$&            9&        \cite{sd} \\
\end{tabular}
\end{table}

\widetext
\squeezetable
\begin{table}
\caption{ A summary of the estimates of the critical parameters for 
$0 \leq N\leq 3$ }
\label{tabella2}
\begin{tabular}{cllll}
$N$ & Method and Ref. & $\beta_c$ & $\gamma$&  $\nu$ \\
\hline
0 &HTE  sc unbiased\cite{mac} &0.2134987(10)& 1.16193(10) &  \\
 &HTE  sc unbiased  &0.213497(6)& 1.161(2) & 0.592(2) \\
 &HTE  sc $\theta$-biased &0.213493(3) & $1.1594(8) $&    $0.5878(6)$\\
 &MonteCarlo sc  \cite{lms}&0.2134969(10) &     $$  & 0.5877(6)  \\
 &MonteCarlo sc  \cite{ca}& 0.213492(1) &1.1575(6)      &   \\
 &HTE  bcc unbiased &0.153131(2)   &   1.1612(8)&    $0.591(2)$\\
 &HTE  bcc $\theta$-biased &0.153128(3) &   1.1582(8)&    $0.5879(6)$\\
 &R.G. FD perturb. \cite{mur}& & $1.1569(8)$&    $0.5872(8) $\\
 &R.G. $\epsilon-$expansion\cite{zinn}& &$1.157(3)$& $0.5880(15)$\\
\hline
1 &HTE  sc unbiased &0.221663(9) & $1.244(3) $&    $0.634(2)$\\
 &HTE sc $\theta$-biased &\big(0.2216544(3)\big)&$1.2388(10) $& $0.6315(8)$\\
 &MonteCarlo sc  \cite{fer}&0.2216595(26)&  & 0.6289(8)  \\
 &MonteCarlo sc \cite{blh}&0.2216544(3)  &1.237(2)    & 0.6301(8)    \\
 &HTE  bcc\cite{fisher}  &   &   1.2395(4)&    $0.632(1)$\\
 &HTE  bcc\cite{nr90}  &   &   1.237(2)&    $0.6300(15)$\\
 &HTE  bcc unbiased &0.157379(2)   &   1.243(2)&    $0.634(2)$\\
 &HTE  bcc $\theta$-biased &\big(0.157373(2)\big) & 1.2384(6)& $0.6308(5)$\\
 &R.G. FD perturb. \cite{mur}& & $1.2378(12)$&    $0.6301(10) $\\
 &R.G. $\epsilon-$expansion\cite{zinn}& &$1.2390(25)$& $0.6310(15)$\\
\hline
2& Experiment\cite{ahl} & & & 0.6705(6)\\
 &HTE  sc unbiased &0.45419(3) & $1.327(4) $&    $0.677(3)$\\
 &HTE  sc $\theta$-biased &0.45419(3) & $1.325(3) $& $0.675(2)$\\
 &MonteCarlo sc  \cite{has}&0.45420(2)&     $1.308(16)$  & 0.662(7)  \\
 &MonteCarlo sc \cite{jan}&0.4542(1) &    $1.316(5) $& $0.670(7)$   \\
&MonteCarlo sc \cite{bal}&0.454165(4) &    $1.319(2) $& $0.672(1)$   \\
 &HTE  bcc unbiased &0.320428(3)   &   1.322(3)&    $0.674(2)$\\
 &HTE  bcc $\theta$-biased &0.320427(3)   &   1.322(3)&    $0.674(2)$\\
 &HTE  fcc \cite{fere}&0.2075(1)  &    $1.323(15) $&    $0.670(7)$\\
 &R.G. FD perturb. \cite{mur}& & $1.318(2)$&    $0.6715(15) $\\
 &R.G. $\epsilon-$expansion\cite{zinn}& &$1.315(7)$& $0.671(5)$\\
\hline
3 &HTE  sc unbiased &0.69303(3) & $1.404(4) $& $0.715(3)$\\
 &HTE  sc $\theta$-biased &0.69305(4) & $1.406(3) $&    $0.716(2)$\\
&MonteCarlo sc \cite{che}&0.693035(37) & $1.3896(70) $& 0.7036(23)\\
&MonteCarlo sc \cite{bal}&0.693002(12) &    $1.399(2) $& $0.7128(14)$   \\
&HTE  bcc unbiased &0.486805(4)  &    1.396(3)&    $0.711(2)$\\
 &HTE  bcc $\theta$-biased &0.486820(4)   &   1.402(3)&    $0.714(2)$\\
&MonteCarlo bcc \cite{che}&0.486798(12)& $1.385(10) $& $0.7059(37)$ \\
&HTE  fcc \cite{mckdom}   &0.3149(6)  &  $1.40(3) $& 0.72(1)   \\
&R.G. FD perturb. \cite{mur}& & $1.3926(26)$&    $0.7096(16)$\\
&R.G. $\epsilon-$expansion\cite{zinn}& &$1.39(1)$& $0.710(7)$\\
\end{tabular}
\end{table}

\widetext
\squeezetable
\begin{table}
\caption{ A summary of the estimates of the critical parameters for 
$4 \leq N\leq 12$  }
\label{tabella3}
\begin{tabular}{cllll}
$N$ & Method and Ref. & $\beta_c$ & $\gamma$&  $\nu$ \\
\hline
4&HTE  sc  unbiased&0.93589(6) & $1.474(4) $&    $0.750(3)$\\
 &HTE  sc $\theta$-biased &0.93600(4) & 1.491(4)&    $0.759(3)$\\
 &MonteCarlo sc \cite{kan} &0.9360(1) &      $1.477(18) $& 0.7479(90)\\
&MonteCarlo sc \cite{bal}&0.935861(8) &    $1.478(2) $& $0.7525(10)$   \\
 &HTE  bcc  unbiased &0.65531(6) &   $1.461(4) $&    $0.744(3)$\\
 &HTE  bcc $\theta$-biased &0.65542(3)   &   1.484(4)&    $0.756(3)$\\
 &R.G. FD perturb.\cite{mal}& &$1.45(3)$& $0.74(1)$\\
 &R.G. FD perturb.\cite{ant}& & $1.449$&    $0.738$\\
\hline
6&HTE  sc unbiased &1.42859(6) & $1.582(5) $&    $0.804(3)$\\
 &HTE  sc $\theta$-biased &1.42895(6) & $1.614(5) $&    $0.821(3)$\\
 &HTE  bcc  unbiased &0.99613(6) &   $1.566(4) $&    $0.796(3)$\\
 &HTE  bcc $\theta$-biased &0.99644(4)   &   1.608(4)&    $0.819(3)$\\
 &R.G. FD perturb.\cite{ant}& & $1.556$&    $0.790$\\
\hline 
8&HTE  sc unbiased &1.9263(2) & $1.656(5) $&    $0.840(3)$\\
 &HTE  sc $\theta$-biased &1.92705(7) & $1.701(4) $& 0.864(3)\\
 &HTE  bcc unbiased &1.33984(7) &   $1.644(5) $&    $0.833(3)$\\
 &HTE  bcc $\theta$-biased &1.34040(6)   &   1.696(4)&    $0.862(3)$\\
 &R.G. FD perturb. \cite{ant}& & $1.637$&    $0.830$\\
 &1/N expansion \cite{oka}& & $1.6449$&    $0.8355$\\
\hline
10&HTE  sc unbiased &2.4267(2) & $1.712(6) $&    $0.867(4)$\\
 &HTE  sc $\theta$-biased &2.42792(8) & $1.763(4) $&    $0.894(4)$\\
 &HTE  bcc unbiased &1.68509(8) &   $1.699(5) $&    $0.860(4)$\\
 &HTE  bcc $\theta$-biased &1.68586(7) & 1.761(4)&    $0.893(3)$\\
 &R.G. FD perturb. \cite{ant}& & $1.697$&    $0.859$\\
 &1/N expansion \cite{oka}& & $1.7241$&    $0.8731$\\
\hline
12&HTE  sc unbiased &2.9291(3) & $1.759(6) $&    $0.889(4)$\\
 &HTE  sc $\theta$-biased &2.9304(1) & $1.812(5) $&    $0.916(4)$\\
 &HTE  bcc unbiased &2.03130(8) &   $1.741(6) $&    $0.881(4)$\\
 &HTE  bcc $\theta$-biased &2.03230(8)   &   1.808(5)& 0.914(3)    \\
 &R.G. FD perturb. \cite{ant}& & 1.743 &    0.881\\
 &1/N expansion \cite{oka}& & $1.7746$&    $0.8969$\\
\end{tabular}
\end{table}


\begin{references}
\bibitem{nota1} Some impressive recent computations in two dimensions 
 are quoted and briefly commented in Ref.\cite{bc2d}.
\bibitem{re} A computation to order $\beta^{20}$ on the sc lattice
has been announced in Ref.\cite{r1}, but no series have yet 
been published.
\end{references}
\end{document}